\documentclass[runningheads]{llncs}
\usepackage[T1]{fontenc}
\usepackage{graphicx,verbatim}
\usepackage{hyperref}
\usepackage[utf8]{inputenc} 
\usepackage{url}            
\usepackage{booktabs}       
\usepackage{amsfonts}       
\usepackage{nicefrac}       
\usepackage{microtype}      
\usepackage{xcolor}         
\usepackage{xspace}
\usepackage{tcolorbox}
\usepackage{pgf-pie}
\usepackage{amsmath}
\usepackage{subcaption}
\usepackage{array}
\usepackage{xcolor} 
\usepackage{amssymb}
\usepackage{pifont}
\usepackage{arydshln}
\usepackage{appendix}

\newcommand{\dataset}{{\textsc{Open-PMC}}\xspace}
\newcommand{\datasetpp}{{\textsc{Open-PMC++}}\xspace}

\newcommand{\biomedclip}{\mbox{{\textsc{PMC-15M}}}\xspace}
\newcommand{\biomedica}{\mbox{{\textsc{Biomedica}}}\xspace}

\newcommand{\arash}[1]{{\color{red}[AA: #1]}}

\definecolor{darkgreen}{rgb}{0.0, 0.5, 0.0} 
\definecolor{darkred}{rgb}{0.5, 0.0, 0.0} 

\def\beqa{\begin{eqnarray}}
\def\eeqa{\end{eqnarray}}
\def\beqann{\begin{eqnarray*}}
\def\eeqann{\end{eqnarray*}}

\begin{document}
\title{Advancing Medical Representation Learning Through High-Quality Data}
%

\author{
    Negin Baghbanzadeh\textsuperscript{1,2}\textsuperscript{*} \and
    Adibvafa Fallahpour\textsuperscript{2,3,5}\textsuperscript{*} \and
    Yasaman Parhizkar\textsuperscript{1,2}\textsuperscript{*} \and \\
    Franklin Ogidi\textsuperscript{2} \and
    Shuvendu Roy\textsuperscript{2,4} \and
    Sajad Ashkezari\textsuperscript{1,2} \and \\
    Vahid Reza Khazaie\textsuperscript{2} \and
    Michael Colacci\textsuperscript{3} \and
    Ali Etemad\textsuperscript{4} \and \\
    Arash Afkanpour\textsuperscript{2}\textsuperscript{$\dagger$} \and
    Elham Dolatabadi\textsuperscript{1,2}\textsuperscript{$\dagger$}
}

\authorrunning{Baghbanzadeh et al.}

\institute{
    \textsuperscript{1}York University \\
    \textsuperscript{2}Vector Institute \\
    \textsuperscript{3}University of Toronto \\
    \textsuperscript{4}Queen's University \\
    \textsuperscript{5}University Health Network
}

\renewcommand{\thefootnote}{}
\footnotetext{\textsuperscript{*} Equal Contribution}
\footnotetext{\textsuperscript{$\dagger$} Equal Advising, arash.afkanpour@vectorinstitute.ai, edolatab@yorku.ca}

\maketitle              
\begin{abstract}
Despite the growing scale of medical Vision-Language datasets, the impact of dataset quality on model performance remains under-explored. We introduce \dataset, a high-quality medical dataset from PubMed Central, containing 2.2 million image-text pairs, enriched with image modality annotations, subfigures, and summarized in-text references. Notably, the in-text references provide richer medical context, extending beyond the abstract information typically found in captions. Through extensive experiments, we benchmark \dataset against larger datasets across retrieval and zero-shot classification tasks. Our results show that dataset quality-not just size-drives significant performance gains. We complement our benchmark with an in-depth analysis of feature representation. Our findings highlight the crucial role of data curation quality in advancing multimodal medical AI. We release \dataset, along with the trained models and our codebase.


\end{abstract}
\section{Introduction}

In the general domain, Vision-Language (VL) modeling has leveraged massive-scale unlabeled image-text pairs to learn useful representations for a wide range of downstream tasks \cite{radford2021learning,jia2021scaling}. The medical domain has also seen a surge in efforts to curate extensive medical image-text datasets, often relying on automated crawling of scientific articles \cite{zhang2023biomedclip,lin2023pmc,pelka2018radiology,lozano2025biomedica}. However, the amount of available data in the medical domain remains significantly smaller compared to the general domain. While increasing the volume of data is one way to train high-performance models, improving data quality remains an under-explored yet promising direction to enhance model performance.

Recent studies \cite{nguyen2022quality} highlight that high-quality datasets can mitigate pretraining limitations and enhance model performance, as evidenced in recent work from DeepSeek \cite{liu2024deepseek}. Yet, in medical AI, automated extraction of figures and corresponding captions from scientific literature introduces quality challenges. Unlike structured medical reports, such as radiology reports that provide detailed anatomical descriptions, figures in scientific articles are often compound images with captions that are brief or lacking essential clinical context. For example, a dermatology report may provide a detailed description of lesion texture and color variation while a figure caption in a scientific article may simply state 'Example of skin lesion', providing little context for pretraining medical VL models.

In this paper, we investigate how the trade-off between \emph{curation quality and data quantity} in the medical domain impacts learned representations.  To address this question, we introduce \dataset, a carefully curated image-text medical dataset, and conduct a comprehensive set of experiments. Specifically, we investigate how image decomposition—where compound figures are replaced with subfigures—and contextually enriched captions improve representation learning. Our contributions are summarized as follows:


\begin{enumerate}
    \item We present \dataset, a high-quality dataset extracted from PubMed Central articles (\href{https://www.ncbi.nlm.nih.gov/pmc/tools/ftp/#indart}{PMC's Open Access Subset}), consisting of 2.2 million paired image-text pairs. Each pair includes \textit{subfigures} as images and \textit{captions}, \textit{subcaptions}, and \textit{summarized in-text references} as text. In addition, all pairs are annotated with imaging modalities, including Radiology, Microscopy, and Visible Light Photography (VLP).
   
    \item Through extensive experiments across three imaging modalities, along with ablation studies, we demonstrate that careful dataset curation improves representation learning, leading to better downstream performance.
    
    
    \item We publicly release \dataset along with the trained models and our codebase, providing the research community with valuable resources for advancing AI in medical imaging.
\end{enumerate}

\section{Related Work}


General-domain multimodal models \cite{radford2021learning,jia2021scaling,girdhar2023imagebind} owe their success to large-scale datasets, sparking interest in curating medical multimodal datasets, which are typically sourced from PMC articles. ROCO \cite{pelka2018radiology} is an open-access dataset which contains approximately 80,000 radiology and 6,000 non-radiology images along with their captions, keywords and other metadata. Lin et al. introduced PMC-OA \cite{lin2023medical} with 1.6 million image-text pairs. They provided a pipeline for extracting image-text pairs to minimize human involvement. We adapt and expand their pipeline for curating our dataset. More recently, \biomedclip \cite{zhang2023biomedclip} was introduced as a large-scale image-text dataset with 15 million pairs. At the time of writing this paper, however, the dataset has not been released publicly. \biomedica~\cite{lozano2025biomedica} is the most recent dataset comprising of 24 million image-text pairs. Using a combination of pretrained image encoders, clustering, and expert annotations they categorized the images with global and local taxonomies, providing modality information for images. Both PMC-15M and Biomedica contain a significant number of non-medical images (plots, charts, etc.). If used for training, this data could hinder model performance in medical applications. In contrast, we have only included high-quality medical images in our dataset. In addition, both \biomedclip and \biomedica contain raw, compound images. Our work builds on these efforts by focusing on high-quality medical images, exploring impact of image decomposition and contextual text augmentation on VL model performance. Table~\ref{tab:dataset_comparison} provides a comparative overview of existing image-text datasets, highlighting differences in size and key features.


\begin{table}[t!]
    \centering
    \caption{Comparison of medical paired image-text datasets.}
    \label{tab:dataset_comparison}
    \scriptsize
    \renewcommand{\arraystretch}{1.0} 
    \setlength{\tabcolsep}{4pt} 
    \resizebox{\textwidth}{!}{
    \begin{tabular}{l c c c c c c c}
    \toprule
    \textbf{} & \textbf{Size (M)} & \textbf{In-text} & \textbf{Summ} & \textbf{} & \textbf{Medical} & \textbf{} & \textbf{Open} \\
    \textbf{Dataset} & \textbf{} & \textbf{Ref} & \textbf{Refs} & \textbf{Subfigures} & \textbf{Only} & \textbf{Modality} & \textbf{Access} \\
    \midrule
    ROCO~\cite{pelka2018radiology}           & 0.08  & {\color{darkred}\ding{55}} & {\color{darkred}\ding{55}} & {\color{darkgreen}\ding{51}}  & {\color{darkgreen}\ding{51}}  & {\color{darkgreen}\ding{51}}  & {\color{darkgreen}\ding{51}}  \\
    PMC-OA~\cite{lin2023medical}        & 1.6   & {\color{darkred}\ding{55}} & {\color{darkred}\ding{55}} & {\color{darkgreen}\ding{51}}  & {\color{darkgreen}\ding{51}}  & {\color{darkred}\ding{55}} & {\color{darkgreen}\ding{51}}  \\
    \biomedclip~\cite{zhang2023biomedclip}     & 15  & {\color{darkred}\ding{55}} & {\color{darkred}\ding{55}} & {\color{darkred}\ding{55}}  & {\color{darkred}\ding{55}}  & {\color{darkred}\ding{55}}  & {\color{darkred}\ding{55}}  \\
    \biomedica~\cite{lozano2025biomedica}      & 24  & {\color{darkgreen}\ding{51}} & {\color{darkred}\ding{55}} & {\color{darkred}\ding{55}}  & {\color{darkred}\ding{55}}  & {\color{darkgreen}\ding{51}}  & {\color{darkgreen}\ding{51}}  \\
    \midrule
    \textbf{\dataset}  & 2.2   & {\color{darkgreen}\ding{51}} & {\color{darkgreen}\ding{51}} & {\color{darkgreen}\ding{51}}  & {\color{darkgreen}\ding{51}}  & {\color{darkgreen}\ding{51}}  & {\color{darkgreen}\ding{51}}  \\
    \bottomrule
    \end{tabular}
    }
\end{table}

\section{\dataset: Curation and Processing}
\begin{figure}[tb]
\centering
\includegraphics[width=0.95\textwidth]{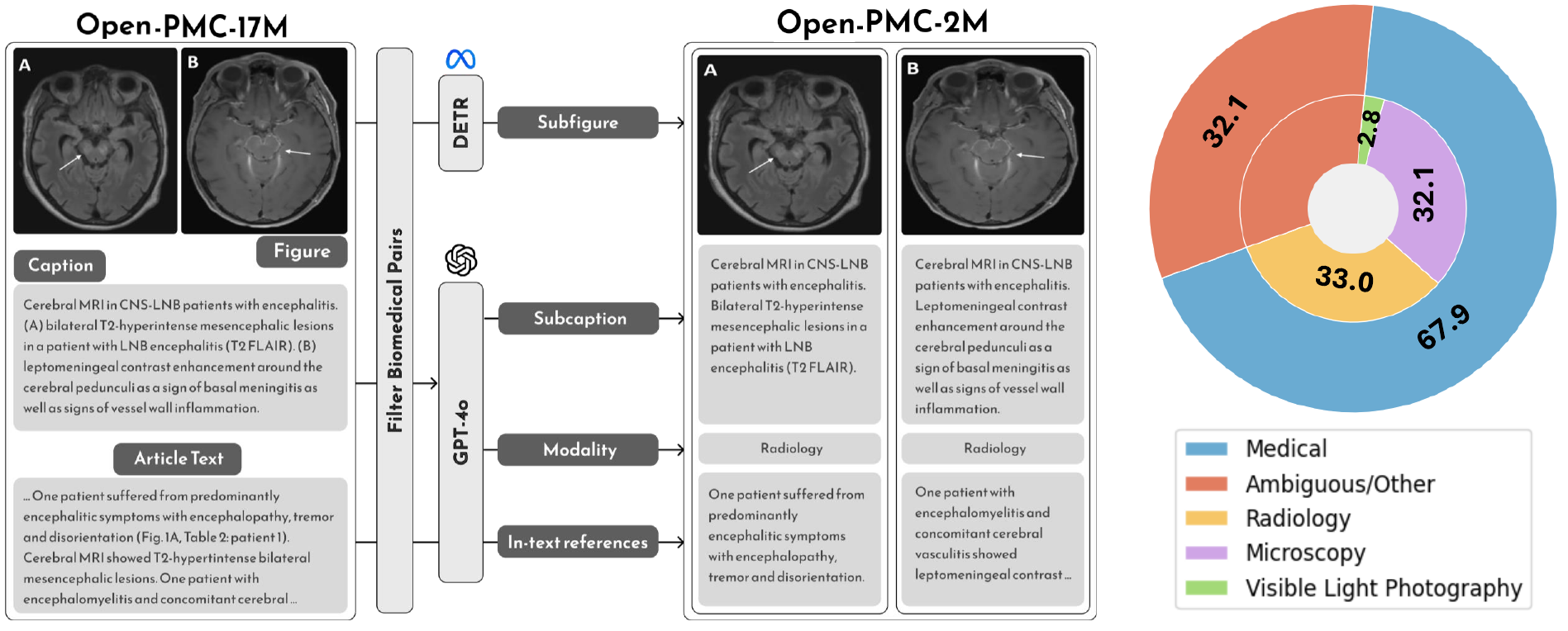}
\caption{\textbf{Left} \dataset-17M comprises 16.7 million image-caption pairs, which undergo rigorous quality curation to produce \dataset, including 2.2 million image-text pairs; images are medical subfigures, and texts are captions enriched with both the actual and summarized content of in-text references. \textbf{Right} The distribution (\%) of each medical image modality within \dataset.}
\label{fig:overall-figure}
\end{figure}

\dataset (Fig.~\ref{fig:overall-figure}) is a fine-grained and open-source dataset of 2.2 million high-quality medical image-text pairs. Each example includes: (1) a medical image (subfigure) extracted from an article, (2) its corresponding caption, (3) an in-text reference from the article body, (4) a \emph{summary} of this reference, and (5) the medical \emph{modality} of the image. To construct \dataset, we extended the PMC-OA pipeline \cite{lin2023pmc}, integrating additional processing steps, including in-text reference extraction and image modality classification using GPT-4o \cite{hurst2024gpt}.

\subsection{Data Collection and Preprocessing}
Our pipeline leveraged \textit{Build-PMC-OA} \cite{lin2023pmc}, processing over four million open-access articles from \href{https://www.ncbi.nlm.nih.gov/pmc/tools/ftp/#indart}{PMC's Open Access Subset} (as of June 18, 2024). We extracted figures, captions, and in-text references, employing XML parsing and regular expressions to link figure-caption pairs with their provenance (PMID and PMC-ID).

\subsubsection{Quality Control and Filtering} We applied a multi-step filtering process, first removing articles with incorrectly formatted XML, missing captions, or syntax errors, yielding $16.7$ million image-caption pairs, \dataset-17M. We then excluded pairs without predefined medical keywords, reducing the dataset to $880,294$ pairs.

\subsubsection{Compound Image Decomposition} We used a DEtection TRansformer (DETR)-based model \cite{carion2020end} trained on a subset of the MedICaT dataset \cite{subramanian2020medicat} to decompose compound images into $3,929,247$ single subfigures. Following decomposition, we used a ResNet-101 model trained on the DocFigure dataset \cite{jobin2019docfigure} to classify images and filter out non-medical samples. This process resulted in a final dataset of 2.2 million high-quality medical images, retaining only figures with high confidence scores in the "Medical" category.

\subsubsection{Caption Segmentation and Alignment} We used GPT-4o to segment full captions into subcaptions for each subfigure. To align captions with subfigures, we leveraged object localization (YOLOv3) and recognition (ResNet-152) from Exsclaim \cite{schwenker2021exsclaim,redmon2018yolov3incrementalimprovement,he2015deepresiduallearningimage} to detect labels and match them to subcaptions. For subfigures without identifiable labels, we assigned the entire original caption to preserve textual context.

\subsection{Textual Augmentation and Contextualization} 
\subsubsection{In-text Reference Extraction and Summarization} We extracted paragraphs referencing a figure in the article using XML cross-reference tags and segment them into individual sentences using regular expressions. However, in-text references often span multiple paragraphs, exceeding the model’s processing limits, and relevant details may appear outside the sentences that directly reference a target subfigure. Moreover, for compound figures, determining which references correspond to specific subfigures is non-trivial. To address these challenges, we used GPT-4o-mini to generate focused summaries, distilling the most relevant contextual information while preserving critical details. 

\subsubsection{Image Modality Assignment}
We used GPT-4o-mini to extract image modality information from captions, categorizing images as diagnostic or non-diagnostic. Diagnostic images were further classified into Radiology (e.g., X-ray, Ultrasound, CT, or MRI), VLP (e.g., Dermatology, Endoscopy), and Microscopy. To validate accuracy, three reviewers independently assessed 1,000 randomly selected images (one per modality), achieving an overall 87\% agreement with the automated classification.

\section{Experiments}
\label{sec:experiments}
We perform an extensive evaluation of \dataset for medical representation learning by training encoders, referred to as VL models, via contrastive learning, which has gained widespread popularity and ubiquity \cite{radford2021learning,zhao2023clip,roy2024benchmarking}. We compare the performance of these models across a wide range of downstream tasks against models trained on similar datasets, some larger than \dataset (specifically, \biomedclip \cite{zhang2023biomedclip}, and \biomedica \cite{lozano2025biomedica}). Since our primary focus in this paper is on the dataset rather than modeling, we refer to datasets rather than model names across all experiments. We conduct our experiments using the \mbox{\textbf{mmlearn}} multimodal learning framework\footnote{ \href{https://github.com/VectorInstitute/mmlearn}{https://github.com/VectorInstitute/mmlearn}}. The code and experimental setup are available at \href{https://github.com/vectorInstitute/pmc-data-extraction}{https://github.com/vectorInstitute/pmc-data-extraction}. 

\begin{figure}[t!]
\centering
\includegraphics[width=\textwidth]{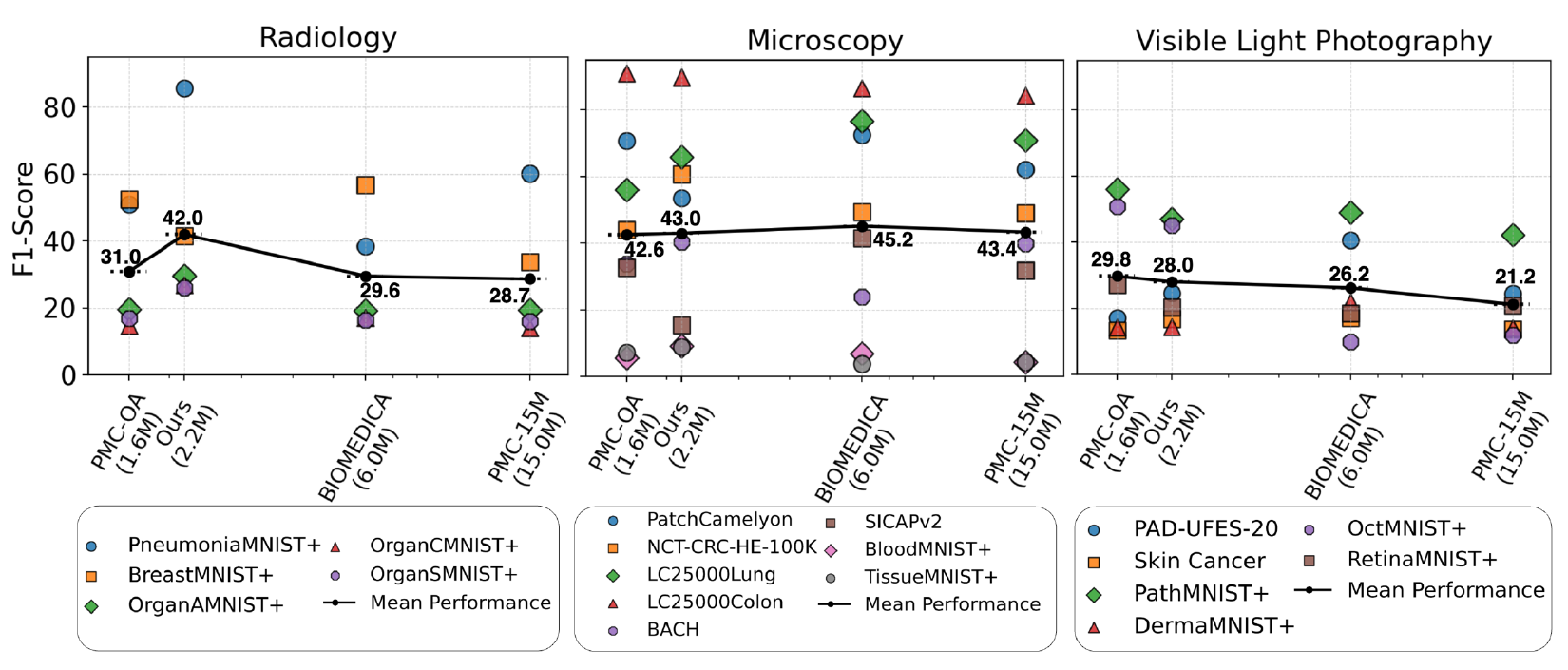}
\caption{Zero-shot classification F1-scores across different VL models trained on datasets of varying sizes, evaluated on downstream tasks split by image modality. Each marker represents performance on an individual task, while the solid line indicates the mean performance across all tasks. \textit{Ours} indicates \dataset.}
\label{fig:zero-shot-plot}
\end{figure}

\subsection{Setup}
\subsubsection{Pretraining} All encoders are trained using a vanilla contrastive loss to align vision and text representations. Our initial encoders comprise PubMedBERT \cite{pubmedbert} as the text encoder and a ViT-B/16 transformer \cite{dosovitskiy2020image}, pretrained on ImageNet, as the vision encoder.
For a fair comparison, we train the same encoder architecture on four datasets: PMC-OA \cite{lin2023pmc}, ROCO \cite{pelka2018radiology}, \dataset, and \dataset without in-text (w/o in-text) references. The \dataset encoders are trained for 64 epochs, while training durations for other datasets are adjusted to ensure all models train on the same total number of examples. For each encoder, the best-performing checkpoint is selected based on validation retrieval performance. Models trained on \biomedclip (BiomedCLIP) and \biomedica ($\text{BMCA-CLIP}_{\text{CF}}$) were downloaded directly from the corresponding HuggingFace pages. 

\begin{table}[htb]
\centering
\caption{Retrieval performance (Recall@200) of VL models. The last two columns, Average Recall (AR) and Mean Reciprocal Rank (MRR) aggregate the results across all tasks. Highest performance values are in bold, second-best are underlined.}
\label{tab:retrieval-results}
\small
\renewcommand{\arraystretch}{1.2}
\resizebox{\textwidth}{!}{
\begin{tabular}{lccc|ccc|cc}
\toprule
 & \multicolumn{3}{c|}{\textbf{Text-to-Image}} & \multicolumn{3}{c|}{\textbf{Image-to-Text}} & \multicolumn{2}{c}{\textbf{Summary}} \\
\textbf{Model} & \textbf{MIMIC-CXR} & \textbf{Quilt} & \textbf{DeepEyeNet} & \textbf{MIMIC-CXR} & \textbf{Quilt} & \textbf{DeepEyeNet} & \textbf{AR} & \textbf{MRR} \\
\midrule
ROCO & 0.080 & 0.024 & 0.079 & 0.084 & 0.023 & 0.108 & 0.066 & 0.208 \\
PMC-OA & 0.139 & 0.142 & \underline{0.152} & 0.152 & 0.149 & \underline{0.157} & 0.149 & 0.361 \\
\biomedclip & \underline{0.162} & \underline{0.186} & 0.147 & \underline{0.185} & \underline{0.166} & \textbf{0.162} & \underline{0.168} & \underline{0.556} \\
\biomedica & 0.094 & \textbf{0.195} & 0.145 & 0.076 & \textbf{0.169} & {0.155} & 0.139 & 0.506 \\
\midrule
\dataset & \textbf{0.170} & 0.166 & \textbf{0.183} & \textbf{0.189} & 0.162 & 0.147 & \textbf{0.170} & \textbf{0.653} \\
\bottomrule
\end{tabular}
}
\end{table}

\subsubsection{Downstream Tasks} We evaluate encoders on retrieval and zero-shot classification tasks. Retrieval includes both image-to-text (I2T) and text-to-image (T2I) retrieval on three benchmark datasets: Quilt \cite{ikezogwo2024quilt} (microscopy), MIMIC-CXR \cite{johnson2019mimic} (radiology), and DeepEyeNet \cite{huang2021deepopht} (VLP). For classification, we conduct zero-shot evaluations across radiology (5 tasks), microscopy (8 tasks), and VLP (6 tasks).


\subsection{Findings}
\subsubsection{Data Quality over Quantity}
Despite \dataset being much smaller (3 to 7 fold) than \biomedica and \biomedclip, the model trained on \dataset achieves performance that is not only comparable but also superior in certain tasks. Fig.~\ref{fig:zero-shot-plot} presents zero-shot classification results, excluding models trained on ROCO due to their consistently low performance. The models trained on \dataset achieve the highest mean performance in radiology, as measured by F1-score, and the least performance variability in VLP. For microscopy classification, it underperforms by only 4.87\% relative to the best model, with a confidence range of 2.87\% to 6.87\%, incorporating an estimated standard deviation of 2\%.

Retrieval results are shown in Table~\ref{tab:retrieval-results}. \dataset achieves the best Average Recall (AR) and Mean Reciprocal Rank (MRR) among the datasets while being smaller than \biomedclip and \biomedica, underscoring the crucial role of data quality in the medical domain. Similar to the classification task, \dataset consistently delivers the highest recall performance in radiology and the best score for T2I retrieval on DeepEyeNet (VLP).

\subsubsection{Distinct Representations vs. Prior Medical Datasets}
\begin{figure}[tb]
\centering
\includegraphics[width=0.95\textwidth]{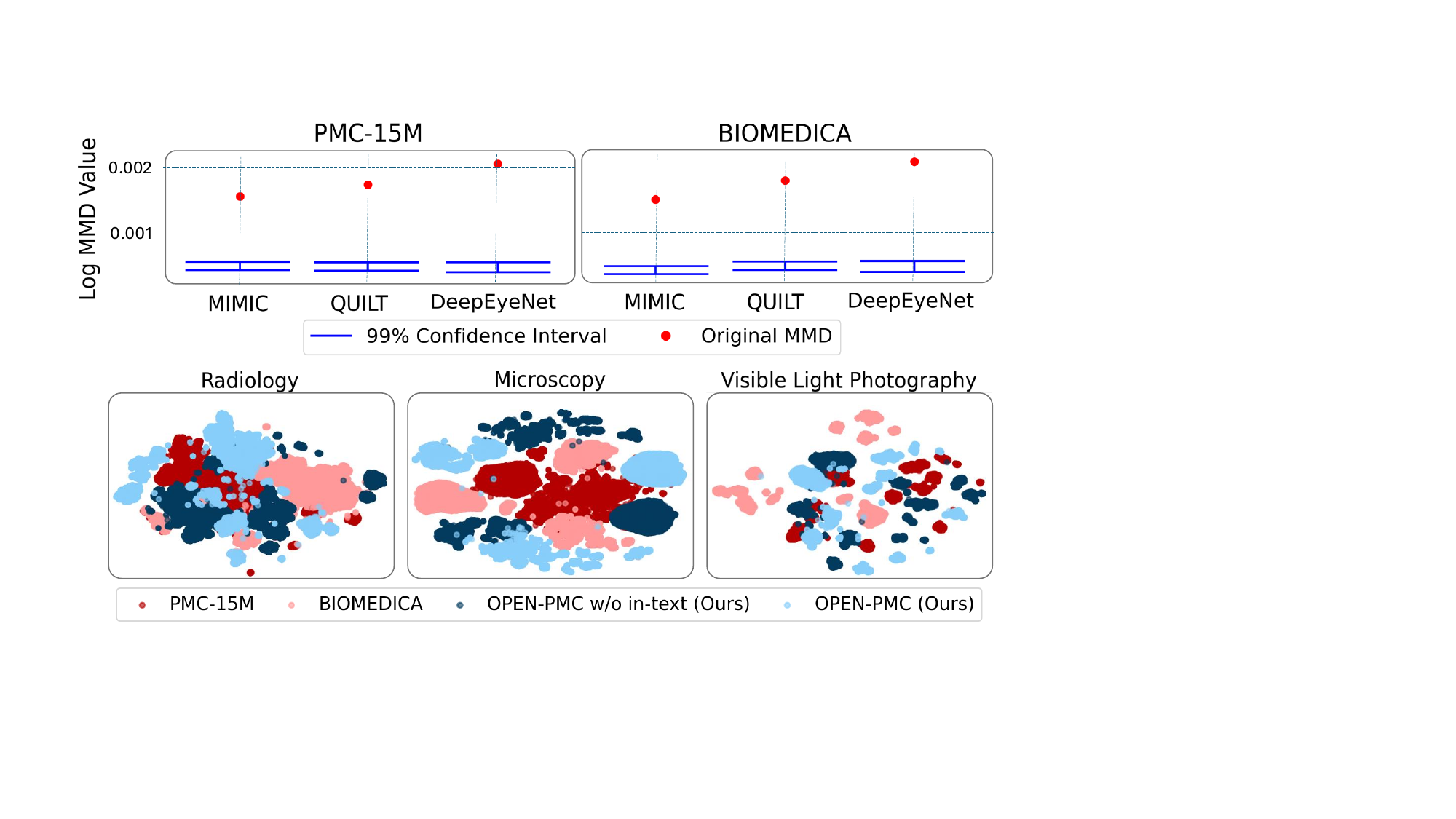}
\caption{Comparison of representation spaces of different VL models. \textbf{(Top)} MMD values between representations learned from \dataset versus \biomedclip and \biomedica. Red dots indicate observed MMD values, and blue bars are 99\% bootstrap confidence interval of the permutation test. \textbf{(Bottom)} t-SNE visualizations of VL models embeddings, illustrating the structure and separation of the learned representation spaces.}
\label{fig:mmd}
\end{figure}

Leveraging VL models trained on \dataset to learn representations for all three medical imaging modalities as shown in Fig.~\ref{fig:mmd} (bottom three plots), reveals that \dataset produces a distinct latent structure compared to \biomedclip and \biomedica in the 2D t-SNE space.

To quantify these differences, we employ Maximum Mean Discrepancy (MMD) \cite{gretton2012kernel}. Let $D$ denote a dataset (e.g., MIMIC-CXR images), and let $\phi(D)$ and $\psi(D)$ denote the feature embeddings obtained by applying encoders $\phi$ and $\psi$ on $D$. For instance, $\phi$ corresponds to an encoder trained on \dataset, while $\psi$ represents one trained on \biomedica. To test whether the underlying distributions of $\phi(D)$ and $\psi(D)$ are different, we conduct a permutation test on the MMD values computed between their respective representation sets. We repeat this experiment with $D \in \{\text{MIMIC-CXR}, \text{Quilt}, \text{DeepEyeNet}\}$ for encoders trained on \dataset, \biomedclip, and \biomedica.

Our MMD analysis reinforces the t-SNE visualization, revealing a substantial difference in the learned representations. As shown in Fig.~\ref{fig:mmd} (top), the MMD values (red dots) between the two representation sets are significantly larger than the 99\% bootstrap confidence interval of the permutation test. This trend remains consistent across MIMIC-CXR, Quilt, and DeepEyeNet. These results indicate the impact of \dataset in shaping different representations from prior medical datasets.

\begin{table}[t!]
    \centering
    \caption{Zero-shot classification F1-score comparison between VL models trained on compound images and subfigures for radiology. Performance differences are shown in parentheses.}
    \label{tab:compound_image_results}
    \renewcommand{\arraystretch}{2} 
    \resizebox{\textwidth}{!}{
    \begin{tabular}{lccccc}
    \toprule
    {\Large \textbf{Model}} & {\Large \textbf{PneumoniaMNIST+}} & {\Large \textbf{BreastMNIST+}} & {\Large \textbf{OrganAMNIST+}} & {\Large \textbf{OrganCMNIST+}} & {\Large \textbf{OrganSMNIST+}} \\
    \midrule
    {\Large \textbf{Compound}}    & {\Large 63.55} & {\Large 48.07} & {\Large 20.83} & {\Large 18.63} & {\Large 18.60} \\
    {\Large \textbf{Subfigures}}  & {\Large 73.58 \color{darkgreen}(10.03 $\uparrow$)} & {\Large 51.47 \color{darkgreen}(3.40 $\uparrow$)} & {\Large 26.68 \color{darkgreen}(5.85 $\uparrow$)} & {\Large 19.96 \color{darkgreen}(1.33 $\uparrow$)} & {\Large 20.89 \color{darkgreen}(2.29 $\uparrow$)} \\
    \bottomrule
    \end{tabular}
    }
\end{table}

\begin{table}[t!]
\centering
\caption{Retrieval performance (Recall@200) comparison between VL models trained on \dataset with and without in-text reference summaries. Performance differences are shown in parentheses, with green indicating higher retrieval performance for \dataset.}
\label{tab:ablation-retrieval-results}
\renewcommand{\arraystretch}{1.5} 
\setlength{\tabcolsep}{16pt} 
\fontsize{55}{35}\selectfont 
\resizebox{\textwidth}{!}{
\begin{tabular}{lccc|ccc|c}
\toprule
 & \multicolumn{3}{c|}{\textbf{Text-to-Image Retrieval}} & \multicolumn{3}{c|}{\textbf{Image-to-Text Retrieval}} & \textbf{Summary} \\
\textbf{Model} & \textbf{MIMIC-CXR} & \textbf{Quilt} & \textbf{DeepEyeNet} & \textbf{MIMIC-CXR} & \textbf{Quilt} & \textbf{DeepEyeNet} & \textbf{AR} \\
\midrule
\dataset w/o in-text & 0.165 & 0.166 & 0.157 & 0.183 & 0.162 & 0.132 & 0.160 \\
\dataset & \textbf{0.170} \color{darkgreen}(0.005 $\uparrow$) & 0.147 \color{darkred}(0.019 $\downarrow$) & \textbf{0.183} \color{darkgreen}(0.026 $\uparrow$) & \textbf{0.189} \color{darkgreen}(0.006 $\uparrow$) & 0.139 \color{darkred}(0.023 $\downarrow$) & 0.147 \color{darkgreen}(0.015 $\uparrow$) & \textbf{0.162} \color{darkgreen} (0.002 $\uparrow$) \\
\bottomrule
\end{tabular}
}
\end{table}

\subsubsection{Ablation Study}
We hypothesize that \dataset's superior performance in radiology tasks stems from the quality of its figures—particularly the use of subfigures instead of compound images—and the inclusion of summarized in-text references added to the caption, both of which are paired with subfigures for contrastive learning. To test this, we conduct two ablation experiments. Since the decomposition pipeline was pre-trained for radiology, our first experiment focuses on radiology classification. We created an alternative version of our dataset where images remained in their original compound form. This reduced the dataset size to 792,000 pairs. To ensure a fair comparison, we randomly sampled an equal number of subfigures from \dataset to match the size of the pairs. As shown in Table~\ref{tab:compound_image_results}, zero-shot classification performance drops when transitioning from subfigures to compound figures. This is one of the factors that distinguishes \dataset from \biomedclip and \biomedica.

The second experiment examines the impact of in-text summaries on retrieval performance. We compared models trained on \dataset with one trained on \dataset w/o in-text summaries on retrieval tasks (Table~\ref{tab:ablation-retrieval-results}). As expected, \dataset outperforms \dataset w/o in-text overall, confirming that incorporating in-text summaries enhances contextual knowledge, strengthening the connection between images and their textual descriptions.

\section{Conclusion}
Our study highlights the critical role of high quality dataset curation in medical VL learning. By introducing \dataset, we demonstrate that image decomposition and incorporating in-text summaries enhance representation learning beyond dataset scale alone. However, a key limitation is that our image decomposition pipeline was primarily optimized for radiology images, limiting its effectiveness for microscopy and other medical modalities. Moreover, further data quality checks and refinement through uncertainty estimation, outlier detection, and human-in-the-loop validation could enhance dataset reliability. Future work will focus on expanding image decomposition techniques for diverse medical imaging modalities and incorporating more robust data quality assurance methods. 


\clearpage
\bibliographystyle{splncs04}
\bibliography{MICCAI/main}

\clearpage
\appendix
\section{Supplementary Material}
\subsection{GPT-4o Prompts}
We utilized the \textbf{GPT-4o and GPT-4o-mini} models for several tasks. First, GPT-4o used it to \textbf{extract subcaptions} associated with each compound image. Additionally, GPT-4o mini was employed to \textbf{summarize in-text references} related to figures and to \textbf{identify the modality} of the images.
In this section, we describe the prompts used for each task.

\subsubsection{Subcaption Segmentation and Alignment}
We utilized GPT-4o (see sample prompt in the Appendix) to segment full captions into panel-specific subcaptions. For captions that could not be reliably separated, the entire caption was retained for the corresponding single-panel figure. To further refine panel-text alignment, we used object localization (YOLOv3) and recognition (ResNet-152) models from Exsclaim \cite{schwenker2021exsclaim,redmon2018yolov3incrementalimprovement,he2015deepresiduallearningimage} to detect labels within subfigures. These detected labels enable automatic matching between subfigures and their corresponding subcaptions. For subfigures without identifiable labels, we assign the entire original caption to preserve textual context.

\begin{tcolorbox}[
    colback=black!4,
    colframe=black!40,
    title=GPT-4o Prompt for Subcaption Extraction,
    boxrule=1.5pt,
    sharp corners,
]
\small
\textbf{System Prompt} \\
Subfigure labels are letters referring to individual subfigures within a larger figure.
Check if the caption contains explicit subfigure label.
If not, output "NO" and end the generation.
If yes, output "YES", then generate the subcaption of the subfigures according to the caption. \\

The output should use the template: \\
    YES \\
    Subfigure-A: ... \\
    Subfigure-B: ... \\
    ... \\\\
The label should be removed from subcaption. \\

\textbf{User Prompt} \\
Caption: {CAPTION}

\end{tcolorbox}

\subsubsection{In-text References}
In-text references provide crucial contextual information about biomedical images, containing clinical interpretations, methodology descriptions, and result analyses that complement the caption. Using XML cross-reference tags, we extract all paragraphs referencing each figure in the article text and break them down into individual sentences using regular expressions. Here, we face three challenges: in-text references span across multiple paragraphs which often exceeds our context length, relevant information may exist outside sentences directly referencing the target subfigure, and for compound figures, it is difficult to determine which references describe which specific subfigures. To address these issues, we employ GPT-4o-mini using the following prompt to generate focused summaries:

\begin{tcolorbox}[
    colback=black!4,
    colframe=black!40,
    title=GPT-4o-mini Prompt to Summarize In-text References,
    boxrule=1.5pt,
    sharp corners,
]
\small
You are provided with a \textbf{medical text} (\textbf{Context}) and a \textbf{target subfigure} (\textbf{Target Subfigure}). Your task is to extract a comprehensive figure caption from the Context that is specific to the Target Subfigure.

\textbf{Required Elements to Include} (if present in Context):
\begin{itemize}
    \item The \textbf{main finding or observation} shown in the Target Subfigure.
    \item Key \textbf{experimental conditions or methods} relevant to the Target Subfigure.
    \item Any \textbf{relevant information} from the Context necessary to understand the Target Subfigure.
\end{itemize}

\textbf{Critical Requirements:}
\begin{itemize}
    \item Do \textbf{not include information} about other subfigures.
    \item Do \textbf{not use external knowledge}; rely solely on the provided Context.
    \item Do \textbf{not add interpretation or analysis} beyond what is explicitly stated.
    \item Ensure all information is \textbf{explicitly present} in the Context.
    \item Keep the caption concise, \textbf{under 250 words}, while preserving essential details.
    \item Focus on content that is \textbf{concise, relevant, objective, and factual}.
\end{itemize}

\textbf{Output Format:}
\begin{itemize}
    \item \textbf{Target Subfigure}: [subfigure number\_label]  
    \item \textbf{Caption}: [detailed caption following the elements above]
\end{itemize}

\textbf{Context}: \{CONTEXT\} \\
\textbf{Target Subfigure}: \{FIGURE\}

\end{tcolorbox}

\subsubsection{Modality Classification}
After separating the compound images and extracting the subcaptions, we utilized \textbf{GPT-4o-mini} to determine the modality of each individual image. For this task, we provided the subcaption of each image to GPT and instructed it to classify the image into one of the following categories: \textbf{Medical}, which includes \textbf{Radiology}, \textbf{Microscopy}, and \textbf{Visible Light Photography}, or \textbf{Ambiguous}. The \textbf{Ambiguous} category accounts for cases where captions are either too brief or lack sufficient information to determine the image modality. The following prompt was used to classify the image modality.

\begin{tcolorbox}[
    colback=black!4,
    colframe=black!40,
    title=GPT-4o-mini Prompt for Image Modality Classification,
    boxrule=1.5pt,
    sharp corners,
]
\small
You are an AI assistant specialized in biomedical topics. You are provided with a text description of a figure (\textbf{Figure Caption}) from a biomedical research paper.

Your task is to classify the image into \textbf{one of four categories} by following a two-step process:

\textbf{Step 1: Determine if the caption is Diagnostic or Non-Diagnostic}
\begin{itemize}
    \item \textbf{Diagnostic Captions}: These are captions of medical images used to directly \textbf{visualize biological structures, detect abnormalities, or assist in clinical decision-making}. They fall into one of the following categories:
    \begin{itemize}
        \item \textbf{Radiology (R)}: X-rays, CT scans, MRI, PET scans, Ultrasound, Angiography, or similar medical imaging methods.
        \item \textbf{Visible Light Photography (V)}: Clinical photographs of external or internal body structures, such as dermatology images (skin lesions), endoscopic images (internal organs), or ophthalmology images (retinal scans).
        \item \textbf{Microscopy (M)}: Images obtained using a microscope to analyze tissue samples, cells, bacteria, or subcellular structures, including light, electron, fluorescence, and transmission microscopy.
    \end{itemize}
    \item \textbf{Non-Diagnostic Captions}: Captions for images that do not directly aid in medical diagnosis but are used for illustration or explanation. They include:
    \begin{itemize}
        \item \textbf{Statistical or Analytical Figures}: Graphs, bar charts, line plots, and pie charts.
        \item \textbf{Diagrams and Schematics}: Flowcharts, system overviews, block diagrams, and hand-drawn illustrations.
        \item \textbf{Data Representations}: Tables, program listings, gene sequences, and molecular structures.
        \item \textbf{Screenshots or UI Elements}: Software interfaces, algorithm visualizations, or non-medical computational images.
        \item \textbf{Non-Clinical Photographs}: Images of lab equipment, experimental setups, or environmental objects.
    \end{itemize}
\end{itemize}

\textbf{Step 2: If the caption is Diagnostic, classify it into one of the following categories:}
\begin{enumerate}
    \item \textbf{R (Radiology)}: Medical imaging techniques used to visualize internal body structures, such as X-rays, CT scans, MRI, PET, Ultrasound, and Angiography.
    \item \textbf{V (Visible Light Photography)}: Photographic images of external or internal body structures, including dermatology, endoscopy, and ophthalmology images.
    \item \textbf{M (Microscopy)}: Any image captured using a microscope, such as light microscopy, electron microscopy, fluorescence microscopy, or transmission microscopy.
\end{enumerate}

If the caption of the image is \textbf{not diagnostic}, return \textbf{N (Non-Diagnostic)}.

\textbf{Response Format:} Your response must be a \textbf{single letter only}, with no explanations:
\begin{itemize}
    \item \textbf{R} for Radiology
    \item \textbf{V} for Visible Light Photography
    \item \textbf{M} for Microscopy
    \item \textbf{N} for Non-Diagnostic
\end{itemize}

\textbf{Caption of the image is:} \{CAPTION\}

\end{tcolorbox}

\subsection{Results}
In this section we will go through more detailed results on the zero-shot retrieval and zero-shot classification results.

\subsubsection{Zero-shot Retrieval}
Tables~\ref{tab:retrieval-results-10} and~\ref{tab:retrieval-results-50} present the zero-shot retrieval performance in terms of Recall@10 and Recall@50, respectively.\\

\noindent
\textbf{MIMIC-CXR}:
MIMIC-CXR is a large publicly available dataset containing over 370,000 chest X-ray images associated with more than 220,000 radiographic studies. The dataset is paired with free-text radiology reports, providing valuable information for medical imaging analysis and natural language processing tasks. It is widely used for developing and evaluating models in automated medical image interpretation.

\noindent\textbf{Quilt}:
Quilt is a curated dataset comprising a diverse collection of medical images spanning various modalities, including radiology, microscopy, and visible light photography. It includes detailed annotations and metadata, making it suitable for training and evaluating models in cross-modal retrieval and classification tasks within the biomedical domain.

\noindent\textbf{DeepEyeNet}:
DeepEyeNet is a specialized dataset focused on ophthalmology, containing high-resolution retinal images used for diagnosing eye diseases. It includes images captured through techniques like fundus photography and optical coherence tomography (OCT), along with corresponding diagnostic labels. The dataset is instrumental for developing models for disease detection and visual understanding in ophthalmic imaging.

\begin{table}[htb]
\centering
\caption{Retrieval performance (Recall@10) of VL models across three datasets. Highest values are in \textbf{bold}, second-best are \underline{underlined}.}
\label{tab:retrieval-results-10}
\small
\renewcommand{\arraystretch}{1.3}
\setlength{\tabcolsep}{10pt}
\resizebox{\textwidth}{!}{
\begin{tabular}{lcccccc}
\toprule
\textbf{Model} & \multicolumn{2}{c}{\textbf{MIMIC-CXR}} & \multicolumn{2}{c}{\textbf{Quilt}} & \multicolumn{2}{c}{\textbf{DeepEyeNet}} \\
 & \textbf{T2I} & \textbf{I2T} & \textbf{T2I} & \textbf{I2T} & \textbf{T2I} & \textbf{I2T} \\
\midrule
ROCO        & 0.004 & 0.005 & 0.002 & 0.002 & 0.006 & 0.006 \\
PMC-OA      & 0.010 & 0.014 & 0.016 & 0.020 & 0.017 & \underline{0.026} \\
PMC-15M     & \underline{0.015} & \underline{0.022} & \underline{0.027} & \underline{0.024} & \underline{0.024} & \textbf{0.031} \\
BIOMEDICA   & 0.006 & 0.005 & \textbf{0.041} & \textbf{0.033} & 0.022 & 0.023 \\
\midrule
OPEN-PMC    & \textbf{0.016} & \textbf{0.022} & 0.016 & 0.018 & \textbf{0.024} & 0.024 \\
\bottomrule
\end{tabular}
}
\end{table}

\begin{table}[htb]
\centering
\caption{Retrieval performance (Recall@50) of VL models across three datasets. Highest values are in \textbf{bold}, second-best are \underline{underlined}.}
\label{tab:retrieval-results-50}
\small
\renewcommand{\arraystretch}{1.3}
\setlength{\tabcolsep}{10pt}
\resizebox{\textwidth}{!}{
\begin{tabular}{lcccccc}
\toprule
\textbf{Model} & \multicolumn{2}{c}{\textbf{MIMIC-CXR}} & \multicolumn{2}{c}{\textbf{Quilt}} & \multicolumn{2}{c}{\textbf{DeepEyeNet}} \\
 & \textbf{T2I} & \textbf{I2T} & \textbf{T2I} & \textbf{I2T} & \textbf{T2I} & \textbf{I2T} \\
\midrule
ROCO        & 0.022 & 0.023 & 0.008 & 0.007 & 0.021 & 0.036 \\
PMC-OA      & 0.044 & 0.054 & 0.056 & 0.062 & 0.070 & \underline{0.071} \\
PMC-15M     & \underline{0.055} & \underline{0.067} & \underline{0.082} & \underline{0.070} & \underline{0.074} & \underline{0.074} \\
BIOMEDICA   & \textbf{0.300} & 0.024 & \textbf{0.102} & \textbf{0.084} & 0.067 & \textbf{0.071} \\
\midrule
OPEN-PMC    & 0.056 & \textbf{0.072} & 0.053 & 0.059 & \textbf{0.077} & 0.058 \\
\bottomrule
\end{tabular}
}
\end{table}

\subsubsection{Zero-shot Classification}
Table~\ref{tab:ZSC_F1} presents the zero-shot classification performance.\\

\noindent\textbf{SICAP}: The Prostate Cancer Grade Assessment (SICAP) dataset consists of histopathological images used for prostate cancer grading. It contains high-resolution images annotated with corresponding cancer grades, supporting research in automated cancer diagnosis and classification.

\noindent\textbf{PAD-UFES-20}: This is a dermatological dataset comprising clinical images of skin lesions collected from the Federal University of Espírito Santo (UFES). The dataset includes detailed metadata and diagnostic labels for various skin conditions, making it valuable for developing skin lesion classification models.

\noindent\textbf{Skin Cancer}: This dataset includes clinical images of skin cancer, labeled with corresponding diagnostic categories. It supports research in early detection and classification of skin cancers through deep learning approaches.

\noindent\textbf{PCam (PatchCamelyon)}: The PCam dataset is derived from histopathological scans of lymph node sections, designed for metastasis detection in breast cancer patients. It includes small image patches labeled as either metastatic tissue or normal tissue.

\noindent\textbf{NCT-CRC-HE}: The NCT-CRC-HE dataset comprises histopathological images of colorectal cancer (CRC) with hematoxylin and eosin staining. It includes various tissue types, supporting research in cancer detection and histopathological image analysis.

\noindent\textbf{LC-Lung}: This dataset contains lung cancer pathology images, labeled according to different lung cancer subtypes. It is used for developing models to assist in the classification and diagnosis of lung cancer.

\noindent\textbf{LC-Colon}: Similar to LC-Lung, this dataset includes pathology images specific to colon cancer, annotated with diagnostic labels to aid in automated classification and diagnosis tasks.

\noindent\textbf{BACH}: The Breast Cancer Histology (BACH) dataset contains microscopy images of breast tissue, annotated across four classes: normal, benign, in situ carcinoma, and invasive carcinoma. It is widely used for developing models for automated breast cancer diagnosis.

\noindent\textbf{DermaMNIST+}: Part of the MedMNIST collection, this dataset consists of dermatoscopic images of skin lesions classified into various skin conditions. It is useful for training and evaluating models in skin disease detection.

\noindent\textbf{OCTMNIST+}: This dataset includes optical coherence tomography (OCT) images of retinal tissues, labeled for common retinal diseases such as choroidal neovascularization, diabetic macular edema, and drusen. It supports research in ophthalmic image classification.

\noindent\textbf{PneumoniaMNIST+}: A subset of chest X-ray images labeled for pneumonia detection. It aids in developing models for automated diagnosis of pneumonia from radiographic images.

\noindent\textbf{RetinaMNIST+}: This dataset comprises retinal fundus images labeled for common eye diseases. It is commonly used for training models in automated retinal disease classification.

\noindent\textbf{BreastMNIST+}: Contains ultrasound images of breast tumors, labeled as benign or malignant. It is used to develop classification models for breast cancer detection.

\noindent\textbf{BloodMNIST+}: This dataset consists of microscopic images of blood cells, classified into various cell types. It is used for automated classification tasks in hematology.

\noindent\textbf{TissueMNIST+}: Contains microscopic images of tissue samples from different organs, labeled according to tissue type, supporting histopathological analysis.

\noindent\textbf{PathMNIST+}: Derived from colorectal cancer tissue slides, this dataset includes images labeled with nine different tissue classes, aiding in multi-class classification tasks in pathology.

\noindent\textbf{OrganAMNIST+}: Consists of abdominal CT images labeled with different anatomical organ classes, supporting organ segmentation and classification tasks.

\noindent\textbf{OrganCMNIST+}: Contains coronal CT images of various organs, labeled for organ classification tasks, and is used for research in medical image understanding.

\noindent\textbf{OrganSMNIST+}: Comprises sagittal CT images of multiple organs, annotated for classification, aiding in comprehensive medical imaging analysis.

\begin{table}[h!]
    \caption{Zero-shot classification F1-scores across diverse medical datasets for different models. The highest scores for each dataset are highlighted in \textbf{bold}.}
    \label{tab:ZSC_F1}
    \centering
    \small
    \resizebox{\textwidth}{!}{
    \begin{tabular}{lccccccc}
    \toprule
    \textbf{Model} & {\rotatebox{0}{\textbf{\fontsize{7}{7}\selectfont PneumoniaMNIST+}}} & {\rotatebox{0}{\textbf{\fontsize{7}{7}\selectfont BreastMNIST+}}} & {\rotatebox{0}{\textbf{\fontsize{7}{7}\selectfont OrganAMNIST+}}} & {\rotatebox{0}{\textbf{\fontsize{7}{7}\selectfont OrganCMNIST+}}} & {\rotatebox{0}{\textbf{\fontsize{7}{7}\selectfont OrganSMNIST+}}} \\
    \midrule
    BIOMEDICA        & 38.46 & 56.66 & 19.25 & 17.13 & 16.33 \\
    BioMedCLIP       & 60.13 & 33.76 & 19.40 & 14.12 & 16.00 \\
    PMC-OA           & 50.94 & 52.36 & 19.70 & 14.79 & 16.99 \\
    ROCO             & 38.46 & 25.07 & 15.91 & 8.94 & 10.00 \\
    \dataset         & 50.13 & \textbf{59.65} & 27.95 & 23.23 & 20.03 \\
    \dataset+Inline  & \textbf{85.59} & 41.57 & \textbf{29.72} & \textbf{27.02} & \textbf{26.08} \\
    \end{tabular}
    }
    \resizebox{\textwidth}{!}{
    \begin{tabular}{lcccccc}
    \toprule
    \textbf{Model} & {\rotatebox{0}{\textbf{\fontsize{7}{7}\selectfont PAD-UFES-20}}} & {\rotatebox{0}{\textbf{\fontsize{7}{7}\selectfont Skin Cancer}}} & {\rotatebox{0}{\textbf{\fontsize{7}{7}\selectfont PathMNIST+}}} & {\rotatebox{0}{\textbf{\fontsize{7}{7}\selectfont DermaMNIST+}}} & {\rotatebox{0}{\textbf{\fontsize{7}{7}\selectfont OCTMNIST+}}} & {\rotatebox{0}{\textbf{\fontsize{7}{7}\selectfont RetinaMNIST+}}} \\
    \midrule
    BIOMEDICA        & 40.57 & 17.20 & 49.10 & 21.89 & 10.00 & 18.53 \\
    BioMedCLIP       & 24.41 & 13.62 & 42.27 & 14.07 & 11.87 & 20.82 \\
    PMC-OA           & 17.18 & 13.30 & \textbf{56.03} & 14.29 & \textbf{50.74} & \textbf{27.22} \\
    ROCO             & 7.85 & 4.85 & 18.80 & 5.78 & 19.88 & 2.88 \\
    \dataset           & 21.11 & 13.56 & 49.16 & \textbf{14.60} & 45.27 & 26.12 \\
    \dataset+Inline  & \textbf{24.66} & \textbf{16.79} & 47.01 & 14.34 & 42.12 & 20.37 \\
    \end{tabular}
    }
    \resizebox{\textwidth}{!}{
    \begin{tabular}{lcccccccc}
    \toprule
    \textbf{Model} & {\rotatebox{0}{\textbf{\fontsize{7}{7}\selectfont Sicap}}} & {\rotatebox{0}{\textbf{\fontsize{7}{7}\selectfont PCam}}} & {\rotatebox{0}{\textbf{\fontsize{7}{7}\selectfont NCT-CRC-HE}}} & {\rotatebox{0}{\textbf{\fontsize{7}{7}\selectfont LC-Lung}}} & {\rotatebox{0}{\textbf{\fontsize{7}{7}\selectfont LC-Colon}}} & {\rotatebox{0}{\textbf{\fontsize{7}{7}\selectfont BACH}}} &  {\rotatebox{0}{\textbf{\fontsize{7}{7}\selectfont BloodMNIST+}}} & {\rotatebox{0}{\textbf{\fontsize{7}{7}\selectfont TissueMNIST+}}} \\
    \midrule
    BIOMEDICA        & 41.53 & 72.57 & 49.46 & 76.63 & 86.54 & 23.88 & 6.83 & 3.86 \\
    BioMedCLIP       & 31.80 & 62.17 & 48.98 & \textbf{70.93} & 84.43 & 39.83 & 4.37 & 4.31 \\
    PMC-OA           & \textbf{32.80} & \textbf{70.65} & 43.95 & 56.04 & \textbf{91.05} & 33.75 & 5.57 & 7.17 \\
    ROCO             & 17.50 & 39.89 & 9.43 & 27.70 & 41.30 & 20.91 & 6.67 & 4.87 \\
    \dataset           & 20.71 & 38.96 & 42.88 & 63.97 & 88.38 & \textbf{41.31} & \textbf{10.73} & 6.08 \\
    \dataset+Inline  & 15.54 & 53.50 & \textbf{60.61} & 65.86 & 89.76 & 40.49 & 9.17 & \textbf{8.90} \\
    \bottomrule
    \end{tabular}
    }
\end{table}

\end{document}